\documentclass[epj]{svjour}
\usepackage{graphics}
\usepackage{natbib}
\begin{document}

\title{Efficient atomization of cesium metal in solid helium by low energy (10~$\mu$J) femtosecond pulses}
\titlerunning{Efficient atomization of cesium metal in solid helium...}
\author{Mathieu Melich\thanks{\emph{Present address~:} Institut N\'eel-MCBT, CNRS, Grenoble} \and Jacques Dupont-Roc \and Philippe Jacquier %
}                     

\institute{Laboratoire Kastler Brossel, ENS, UPMC-Paris 6, CNRS ; 24 rue Lhomond, 75005 Paris, France}
\date{Received: date / Revised version: date}
%
\abstract{ %
Metal atoms in solid and liquid helium-4 have attracted some interest either as a way to keep the atoms in a
weakly perturbing matrix, or using them as a probe for the helium host medium. Laser sputtering with
nanosecond pulsed lasers is the most often used method for atom production, resulting however in a
substantial perturbation of the matrix. We show that a much weaker perturbation can be obtained by using
femtosecond laser pulses with energy as low as 10~$\mu$J. As an unexpected benefit, the atomic density
produced is much higher.
} 
\maketitle
\section{Introduction}\label{intro} %
Over the past twenty years, atomic and molecular impurities in superfluid and solid helium-4 have been
extensively studied by spectroscopic methods either in bulk (for a review \cite{tabbert97,moroshkin06b}) or
in clusters \cite{toennies04,stienkemeier06}. Among the motivations of those works, one may cite the
investigation of superfluidity at a microscopic level \cite{hartmann96b,grebenev98}, the reactions of atoms,
molecules, radicals in the helium matrix \cite{lugovoj00,reho01}, investigation of matrix excitations by
optical methods \cite{hartmann96b,ishikawa97,hui00}, measurement of electron EDM \cite{arndt93,moroshkin06b}
or parity violating nuclear anapole moment \cite{bouchiat01}. Several methods (recombination \cite{bauer90},
laser sputtering \cite{fujisaki93,arndt93}, jet \cite{gordon04}) have been developped to introduce atomic
impurities into bulk helium. Laser sputtering has been quickly recognized as the most efficient method to
introduce various atomic species into liquid \cite{fujisaki93,arndt93,beijersbergen93,ishikawa97} and solid
helium \cite{kanorsky94a}. Atomic density about $10^{14}$-$10^{15}$~m$^{-3}$ has been reported
\cite{fujisaki93,arndt95b}. In this method, atomization proceeds in two steps. In the first step, a metal
target situated in condensed helium is sputtered that produces metal grains and clusters embedded in the
medium. These grains are then atomized by a second laser and detected by a third one. In superfluid,
convection flow induced by the sputtering process carries away the atoms from the detection region in a few
milliseconds \cite{fujisaki93}. An improvement was later achieved leading to the detection of atomic Cs in
superfluid helium over about 500~ms \cite{furukawa06}. In solid helium, atoms are more efficiently trapped
and can be observed during several hours \cite{arndt95b} or even a week \cite{melich08}.

Laser sputtering has some drawbacks however. When YAG second harmonic (532~nm) pulses are used, pulse
energies about 1-10~mJ are necessary. This creates transient bubbles in the liquid or melting of the solid
matrix over a macroscopic volume. These energetic events are likely to strongly perturb the helium matrix.
Indeed evidence for the conservation of the global solid orientation is still lacking
\cite{kanorsky98,melich08}.  A first step  may consist in reducing the energy brought by the laser.

In this paper a new method is described making use of amplified femtosecond pulses to atomize cesium grains
or clusters. Due to much shorter pulse duration, much lower energies (10~$\mu$J) are used to obtain similar
or even greater atomic densities than previous methods. To our knowledge, metal atomization with femtosecond
pulses  has been reported only once \cite{furukawa06}, but without details about the atomic densities
obtained. We describe first the experimental set-up used and the atomization process of cesium impurities in
an initially single helium crystal. Atomic densities obtained are then discussed.

\section{Experimental set-up}
\begin{figure}[!hb]
\includegraphics{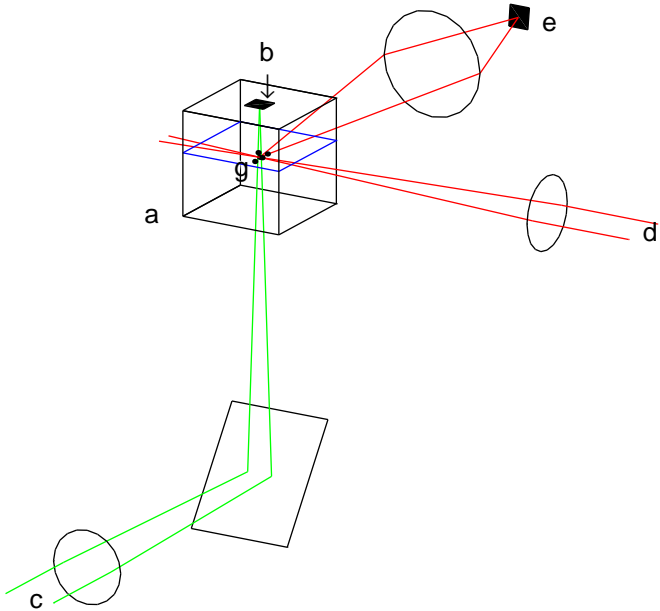}
 \caption{Sketch of the experimental arrangement. (a) Cell inside the cryostat containing solid He at the bottom, (b) Solid cesium target immersed into HeII, (c) Sputtering YAG laser, (d) Amplified Ti:Sa laser for atomization (mode-locked) or in cw mode for detection, (e) fluorescence  detection, (g) cesium metal grains and clusters.} \label{fig1}
\end{figure}

A sketch of the experimental arrangement is given in figure~\ref{fig1}. A 2~cm wide cell is attached to a
helium-4 refrigerator. Its temperature can be regulated  between 1 and 1.8~K. To grow a hcp helium crystal,
liquid helium is pressurized at constant temperature, typically 1.2~K, up to the solidification pressure
(25.4~bar). An electrostatic device \cite{keshishev81} nucleates the solid phase as a small single crystal
which falls on the bottom of the cell. It is then grown by continuously feeding the cell with helium until
the liquid/solid interface reaches the middle of the cell windows. In order not to damage the crystal by the
first step of the atomization process, we put the cesium target in the liquid above the solid. Sputtering is
produced by second harmonic YAG laser pulses (energy 5~mJ, repetition rate 2~Hz). Sub-millimetric cesium
grains, and possibly clusters, fall onto the crystal surface. When their number is sufficient, they are
embedded into the crystal by a further growth.

They lie then 1 or 2~mm below the solid/liquid interface. The next section explains how they are later
atomized by femtosecond laser pulses.

In  order to measure the local density of cesium atoms, laser induced fluorescence is used. A Ti:Sa laser
(\lq d\rq{} in figure 1) can be used in cw or mode-locked regime. In the cw regime, it excites the
fluorescence on the D1 line at 850~nm. The absorption line is shifted with respect to the free atom line due
to the interaction of cesium atoms with the helium matrix \cite{kanorsky95}. The fluorescence light is
collected at right angle, selected by an interference filter centered at 880~nm and its intensity is
measured by a cooled APD (Hamamatsu C4777-01). The use of femtosecond laser pulse for metal sputtering or
machining is now well documented \cite{nolte97} and various kinds of amplified sources have been described.
In the present case, low average power is desirable in order to keep the low temperature sample with a
minimal perturbation. Hence we require a low repetition rate and a pulse energy sufficient to atomize metal
grains a focal distance of 0.2~m, comparable with the cryostat radius. Pulse energy on the order of a few
microjoules is sufficient. Much shorter focal length would allow to reduce this energy, but would require
the focusing lens to be located inside the cryostat, which we wanted to avoid for practical reasons.
Microjoule pulse energies are intermediate between the nanojoule energies delivered by Ti:Sa oscillators and
millijoules currently reached by the chirped pulse amplifiers. Simple lasers are not available for such
pulse energy. Hence we built a simple home made multipass unsaturated amplifier which does not require the
100 femtosecond pulse to be chirped. Non-linear effects in the Ti:Sa crystal are kept below the critical
value for self-focusing if pulse energy does not exceed 40~$\mu$J. The Ti:Sa oscillator produces 5~nJ,
200~fs pulses at the rate of 70~MHz. A Pockels cell selects one pulse every 0.1~s. It enters a 4{\em
f}-cavity where it is amplified 6 times in a 1.5~cm Ti:Sa crystal pumped by two 7~mJ second harmonic Nd-YAG
pulses. The pulse energy is then $6 \pm 2~\mu$J. Another opportunity is to let all oscillator pulses go
through the amplifier. Then trains of about 12 successive pulses are amplified synchronously with the YAG
pump pulses. Other Ti:Sa pulses go through the amplifier unaffected. In the time between successive pulses,
atoms and clusters have already been expelled few microns away but have had no time to recombine. Hence
clusters produced by one pulse may be further atomized by the next ones before being thermalized and
possibly recombine. One expects a more efficient atomization in this multi-pulse regime.

\section{Atomization process inside solid helium}
Once cesium grains are embedded 2~mm below the crystal surface, atomization with the Ti:Sa laser begins. The
wavelength is tuned to 850~nm so that fluorescence of the atoms will be excited. Monitoring the atomization
is done by replacing the APD by a fiber/CCD spectrograph (Ocean Optics 4000). Two detection bands are
defined. One from 845 to 855~nm measures the incident light scattered by cesium grains and clusters. The
other one measures atomic fluorescence from 870 to 890~nm. This LIF detection is specific to cesium
monomers. Dimers and trimers do not fluoresce in this band when excited at 850~nm \cite{bunermann04}. Larger
cesium clusters are not known to fluoresce.

\begin{figure}[!hb]
\scalebox{0.9}{\includegraphics[20,40][300,280]{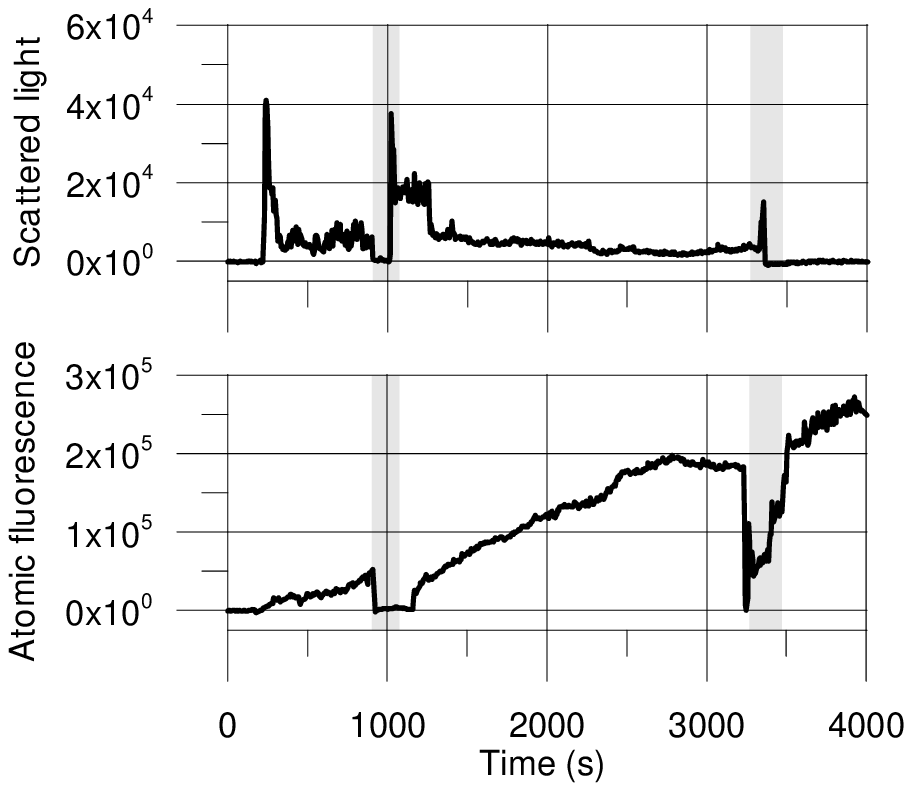}}
 \caption{Atomic fluorescence (lower sub-figure) and incident light scattered by cesium grains and clusters %
(upper sub-figure) during an atomization process at 850~nm. The growth of the atomic fluorescence is
correlated  with the disappearence of the grains scattering the incident light. Gray areas indicate when the
laser focus is moved from one place to a new one, free of pre-existing cesium monomers.} \label{fig2}
\end{figure}

Evolution of both intensities during an atomization process are shown in figure~\ref{fig2}. The Ti:Sa laser
is mode-locked with an averaged intensity 50~mW in the cell, with amplified bursts at a rate of 2~Hz. The
burst contribution to the average intensity is negligible. During the sequence shown, the focal point of the
laser is moved from time to time to explore the region in which clusters are located. This is indicated by
gray areas in figure \ref{fig2}. Intensity of elastically scattered light varies strongly according to the
focal point location, probably in relation with the presence of grains. Keeping the laser focus at such
place and continuing laser irradiation, one observes a slow decrease of the scattered light intensity, while
the atomic fluorescence intensity increases strongly. Note the difference in scale (factor 10) for the two
intensities~: the atomic fluorescence is much stronger. After 20 minutes at this place, the scattered light
has nearly disappeared and the atomic fluorescence does not increase anymore. A straightforward explanation
is that successive laser pulses break metal grains into smaller and smaller pieces, and finally into atoms.
In some cases, no light is scattered, indicating there are no grains while fluorescence does appear after
some time. In these situations, source of atoms are very likely clusters. As we will see now, this
atomization process is indeed efficient.

\section{Estimate of the atomic density}

Once atomic fluorescence intensity is strong enough, a measurement of the fluorescence intensity is made to
provide an estimate of the atomic density. The Ti:Sa laser is switched to its cw mode, while keeping its
wavelength at 850~nm. The beam is chopped at 173~Hz and the APD output is measured through a lock-in
amplifier. The detected fluorescence intensity $I_{det}$ is proportional to the mean atomic density
$\bar{\rho}$ in the region where the two beams intersect, to the laser intensity $I_{exc}$ and to the
absorption cross-section $\sigma$ for the D1 line. More precisely, $\bar{\rho}$ is given by
\begin{equation}
 \bar{\rho} = \frac{1}{\eta\,\sigma\,l_{det}\,\kappa}\,\frac{4\,\pi}{\Omega}\,\frac{I_{det}}{I_{exc}} \nonumber
\end{equation}
where $\Omega$ is the solid angle of the collected light (0.12~sr), $\kappa$ is the transmission efficiency
of the detection beam estimated to be 0.38 at 880~nm, $\eta$ is the fluorescence efficiency for the D1 line
measured to be 0.9 at 25~bar \cite{hofer08}. This formula takes into account the fact that the waist of
Ti:Sa beam in the cell (0.3~mm) is smaller than the diameter of the area seen by the APD detector ($l_{det}
= 1.1$~mm). The absorption cross-section $\sigma$ is the resonant cross-section for the D1 line
$\lambda^2/2\pi$ \cite{jackson75,cagnac02}, scaled by the ratio of the natural linewidth (FWHM)
($\Gamma/2\pi = 4.5$~MHz) to the linewidth in the helium matrix ($\Delta\lambda \simeq 10$~nm). This gives
$\sigma \simeq 1.3 \times 10^{-19}$~m$^2$.

For a 50~$\mu$W excitation intensity, we have measured atomic fluorescence intensities up to $8 \times
10^{-11}$~W. According to the formula above, this corresponds to an atomic density about $3.4 \times
10^{18}~\textrm{m}^{-3}$. This is about 4 orders of magnitude higher than what have been reported in solid
helium \cite{arndt95b}. We have obtained repeatedly densities over $10^{17}~\textrm{m}^{-3}$.\\ As already
mentionned the atomic density depends strongly on the number and sizes of the metal grains in the sputtered
region. One can get densities varying by several order of magnitude for apparently similar conditions.
 Note that the atomic density does not vary significantly over the time required for the density measurement.
As already reported, at 1.0~K, atomic density decays only over several days \cite{melich08,melichPhD}.

\section{Conclusion}

We have shown that femtosecond laser pulses with energy as low as 10~$\mu$J are an efficient tool to atomize
metals in solid helium. Such low energy pulses with a repetition rate of a few hertz are suitable for
experiments at low temperatures below 1~K where the cooling power of helium refrigerators drops down. These
pulses can be produced by simple amplifiers without a stretching/compression device. Substantial atomic
densities have been obtained, larger than 10$^{17}$~atom/m$^{3}$. However we have no evidence that
atomization with such low energy pulses still does not damage the crystal order. In a further step we plan
to attempt atomization directly above crystal surface, in the liquid, and to grow the crystal from this \lq
cesium solution\rq{}. This should allow for obtaining good quality doped single crystal.
\begin{acknowledgement}
We are indebeted to B\'eatrice Chatel (LCAR, Toulouse) for the loan of the Ti:Sa laser. This work was
supported by the ANR contract META (ANR-05-BLAN-0084-02).
\end{acknowledgement}

\end{document}